\documentclass[12pt]{article}

\usepackage{graphicx,psfrag,epsf}
\usepackage{enumerate}
\usepackage{natbib}
\usepackage{url} 
\usepackage[T1]{fontenc}
\usepackage[utf8]{inputenc}
\usepackage{amsmath,amsfonts,amssymb,amscd,amsthm,xspace}
\usepackage{caption}
\usepackage[labelformat = empty,position=top]{subcaption}
\usepackage{xypic}
\usepackage[color,matrix,arrow,curve]{xy}
\usepackage{rotating}
\usepackage{tikz}
\usepackage{csvsimple}
\usepackage{adjustbox}
\newcommand{\norm}[1]{\left\lVert#1\right\rVert}
\DeclareMathOperator*{\argmin}{arg\,min}
\newcommand\independent{\protect\mathpalette{\protect\independenT}{\perp}}
\def\independenT#1#2{\mathrel{\rlap{$#1#2$}\mkern2mu{#1#2}}}

\numberwithin{equation}{section}

\newtheorem{theorem}{Theorem}

\begin{document}

\def\spacingset#1{\renewcommand{\baselinestretch}%
{#1}\small\normalsize} \spacingset{1}

  \title{\bf Graphical Model Selection for Gaussian Conditional Random Fields in the Presence of Latent Variables}
  \author{Benjamin Frot\thanks{
    BF gratefully acknowledges the EPSRC and Amazon Web Services. LJ is supported by
    a Wellcome Trust grant (098759/Z/12/Z) and Christ Church, Oxford. GM 
    is funded by the Wellcome Trust grant 100956/Z/13/Z. This work was jointly supervised by LJ and GM.}\hspace{.2cm}\\
    Department of Statistics, University of Oxford\\
    and \\
    Luke Jostins \\
    Wellcome Trust for Human Genetics, University of Oxford\\
    and \\
    Gilean McVean\\
    Department of Statistics, University Of Oxford}
  \maketitle

\bigskip

\begin{abstract}
We consider the problem of learning a conditional Gaussian graphical model in the presence of latent variables.
Building on recent advances in this field, we suggest a method that decomposes the parameters of a conditional
Markov random field into the sum of a sparse and a low-rank matrix. We derive convergence bounds
for this estimator and show that it is well-behaved in the high-dimensional regime
as well as ``sparsistent'' (\emph{i.e.} capable of recovering the graph structure). We then show
how proximal gradient algorithms and semi-definite programming techniques can be employed
to fit the model to thousands of variables.
Through extensive simulations, we illustrate the conditions required for
identifiability and show that there is a wide range of situations in which this model
performs significantly better than its counterparts, for example, by accommodating more latent variables. Finally, the suggested method is applied
to two datasets comprising individual level data on genetic variants and metabolites levels. We show our results
replicate better than alternative approaches and show enriched biological signal.
\end{abstract}

\noindent%
\vfill

\newpage
\spacingset{1.45} 

\section{Introduction}

The task of performing graphical model selection arises in many applications in science and engineering. 
There are several factors that make this problem particularly challenging.
First, it is common that
only a subset of the relevant variables are observed and estimators that do not account
for hidden variables are therefore prone to confounding.
On the other hand, modelling latent variables is itself 
difficult because of identifiability and tractability issues. 
Second, the number of variables being modelled is often greater than the number of samples. 
It is well known that, in such a scaling regime, obtaining a consistent estimator is usually impossible without making further assumptions about the model, \emph{e.g.} sparsity or low-dimensionality. 
Finally, modelling the joint distribution
over all observed variables is not always relevant. 
It is sometimes preferable
to learn a graphical model over a number of variables of interest while conditioning on the rest of the collection.

These problems are encountered in many fields of application. 
In genetics, for example, one might model a gene expression network conditional on the samples’ combinations of DNA variants: the variables of interest are the expression levels, while the DNA variants are included because of their predictive power and capacity to explain some of the observed correlations between genes \citep{Stearns2010}. As genotype is not causally influenced by gene expression levels (i.e. the direction of effect only goes genotype to expression), we would like to model expression levels conditional on genotype. For another example, consider the task of modelling stock returns conditional on sentiment analysis data. The variables that encode sentiment about the stocks have value \citep{Li2014}, but modelling their joint distribution might be difficult and unnecessary, hence the need for conditioning. Moreover, a number of unmeasured variables (\emph{e.g.} energy prices) might impact many stocks and should be modelled for better predictive accuracy \citep{Chandrasekaran2012}.

The problem of learning a Gaussian graphical model in the presence of latent variables was considered by \cite{Chandrasekaran2012}. 
They suggest estimating an inverse covariance matrix which is the sum of a sparse and a low-rank matrix.
Another partial solution to our problem was introduced independently by \cite{Sohn2012} and \cite{Wytock2013} who defined the concept of a \emph{sparse Gaussian conditional random field}: a regularised maximum-likelihood estimator that learns a Gaussian graphical model over a subset of the variables ($X$, say) while conditioning on the remaining variables ($Z$, say).

\cite{Chandrasekaran2012}, \cite{Sohn2012} and \cite{Wytock2013} made significant advances to the problem of model selection in general graphical models, but there exist many situations where we may wish to allow for latent variables and condition on some of those measured. 
Here, we suggest learning a Gaussian 
conditional random field in the presence of latent variables and introduce a novel regularised maximum-likelihood estimator which fits into the ``low-rank plus sparse'' framework \citep{Chandrasekaran2009, Candes2011}. In our setting, inputs (variables in $Z$) are allowed to act on the outputs ($X$) in both a sparse and a low-rank fashion, while the inverse covariance matrix over $X$ is estimated conditional on $Z$ and on the marginalised latent variables. As will be shown later, this approach allows us to correctly recover graphs that are typically denser and with more hidden variables than the ones that can be handled by other methods. 

From both a theoretical and a computational point of view, modelling latent variables with a conditional random field gives rise to a number of complications (\emph{e.g.} the proximal operator is not defined in closed form) that we address in this paper. In particular, we derive
convergence bounds for our estimator and show that under suitable identifiability conditions it is consistent in the high-dimensional regime as well as ``sparsistent'' (\emph{i.e.} capable of recovering the graph structure).
We then show how the alternating direction method of multipliers \citep{Boyd2010} and semi-definite programming techniques can be employed to fit the model to thousands of variables.
Through extensive simulations, we illustrate the conditions required for identifiability and show
that there is a wide range of situations in which this model performs significantly
better than its counterparts.
In order to show how our model behaves in a realistic setting, we apply the present estimator to two
datasets comprising genetic variants and metabolite levels. Both replication and a test statistic constructed using an independent source of validation suggest that our estimates have more biological relevance than the results obtained via other methods.
\section{Problem Statement}
\label{problem_statement}
\subsection{Setup}
\label{setup}
Throughout, we consider $n$ independent, identically distributed realisations of a zero-mean random vector 
$Y \in \mathbb{R}^{m + p + h}$. $Y$ is indexed by disjoint subsets of $\{1, \ldots, m + p + h\}$, denoted $Z, X, H$ and with respective cardinality $m$, $p$ and $h$. They correspond
to the variables we wish to condition on, the variables we wish to model and the hidden
variables. We write $Y_{Z}$ (resp. $Y_X$ and $Y_H$) for the subvector of $Y$ indexed by $Z$ (resp. $X$ and $H$). Our main assumption is that the distribution of $\begin{pmatrix} Y_X\\Y_H \end{pmatrix} \in \mathbb{R}^{p + h}$ conditional on $Y_Z \in \mathbb{R}^{m}$ is
normal and that its mean is a linear combination of the inputs $Y_Z$. More precisely, we assume a Gaussian conditional random field parametrised as follows:
\begin{equation*}\label{gcrf1}
    \begin{pmatrix} Y_{X} \\
    Y_{H}
    \end{pmatrix} | Y_Z \sim \mathcal{N}\left\{-\begin{pmatrix} M^\ast_X& M^\ast_{XH}\\ {M^\ast_{XH}}^T & M^\ast_H \end{pmatrix}^{-1} 
    \begin{pmatrix} {M^{\ast}_{ZX}}^T\\ {M^{\ast}_{ZH}}^T \end{pmatrix} Y_Z, \begin{pmatrix} M^\ast_X& M^\ast_{XH}\\ {M^\ast_{XH}}^T & M^\ast_H \end{pmatrix}^{-1} \right\},
\end{equation*}
where we have used partitioned matrices to show the contributions 
of the observed and hidden variables. Thus, $M_X^\ast \in \mathbb{R}^{p\times p},
M_{ZX}^\ast \in \mathbb{R}^{m \times p}, M_{XH}^\ast \in \mathbb{R}^{p \times h}, \ldots$. The superscript $-^\ast$ is used to indicate that these matrices are nominal parameters of the model, as opposed to estimates.
Note that there are no distributional assumptions about $Y_Z$. 

Finally, we assume that variables indexed by $H$ are unobserved. Accordingly, 
we compute the marginal distribution $Y_X | Y_Z$, which yields
\begin{equation}\label{gcrf}
    Y_{X} | Y_Z \sim \mathcal{N}\left\{
    -\left(S^\ast_X - L^\ast_X\right)^{-1}
    \left( {S^\ast_{ZX}}^T - {L_{ZX}^\ast}^T \right) Y_Z, 
    \left(S^\ast_X - L^\ast_X\right)^{-1} \right\},
\end{equation}
where we have defined $S_X^\ast \triangleq M_X^\ast$, 
$L_X^\ast \triangleq M_{XH}^\ast {M_X^\ast}^{-1} {M_{XH}^\ast}^T$, $S_{ZX}^\ast \triangleq M_{ZX}^\ast$
and $L_{ZX}^\ast \triangleq M_{ZH}^\ast {M_H^\ast}^{-1} {M_{XH}^\ast}^T$. This expression 
follows straightforwardly from the formula for the inverse of a partitioned matrix (the full derivation is given in the supplementary materials). From Equation \eqref{gcrf}, the log-likelihood function can be expressed in terms of the sample covariance matrices $\Sigma_Z^n \triangleq \frac{1}{n} \sum_i ({Y_Z})_i({Y_Z})_i^T , \Sigma_X^n \triangleq \frac{1}{n} \sum_i ({Y_X})_i ({Y_X})_i^T$ and $\Sigma_{ZX}^n \triangleq \frac{1}{n} \sum_i ({Y_Z})_i ({Y_X})_i^T$:
\begin{multline}\label{gen_ll}
    \ell(S_X, L_X, S_{ZX}, L_{ZX}; \Sigma^n_Z, \Sigma^n_X, \Sigma_{ZX}^n) = \log\det \left(S_X - L_X \right) - Tr\left(\Sigma_X^n (S_X - L_X)\right) - \\
    2 Tr\left(\Sigma_{ZX}^n (S_{ZX} - L_{ZX})^T\right) - Tr\left(((S_X - L_X)^{-1}(S_{ZX} - L_{ZX})^T \Sigma_Z^n (S_{ZX} - L_{ZX})\right).
\end{multline}
For clarity, all terms related to a given subset will be dropped from the expression when the subset is empty. For example, whenever $Z = H = \emptyset$ the log-likelihood becomes $\ell(S_X; \Sigma_X) = \log \det S_X - Tr(\Sigma_X^n S_X)$.

Note that our assumption about the Gaussianity of $X, H$ plays an important role in the interpretation of the nominal parameters ($M_X^\ast, M_{XH}^\ast, \ldots$). Under this assumption, it is well known that the structure 
of the conditional Gaussian graphical model (GGM) over $X, H$ can be read-off these matrices directly by looking at the location of their non-zero entries \citep{Lauritzen1996}. Briefly, a 
graphical model is a statistical model defined according to a graph whose nodes are random variables and whose
edges encode conditional independence statements between variables \citep{Lauritzen1996}. Thus, $(M_{X}^\ast)_{i,j} =  (M_{X}^\ast)_{j,i} = 0$ if and only if $X_i \independent X_j | Z, X \setminus \{X_i, X_j\}, H$. Likewise, $(M_{ZH}^\ast)_{i,j} = 0$ if and only if $Z_i \independent H_j | Z \setminus {Z_i}, X, H \setminus \{H_j\}$. Note that since the conditional mean vector is a linear transformation of $Y_Z$, this interpretation of the non-zero entries of $M_{ZH}^\ast$ and $M_{ZX}^\ast$ holds irrespective of $Y_Z$'s distribution.

\subsection{Goal}\label{goal}

In typical applications such as the ones mentioned in the introduction, $S_X^\ast$ is the target. Since it encodes the structure of the graphical model over $X$, recovering $S_X^\ast$ can provide insight into the causal mechanisms underpinning the data but, in general, hidden
variables make it impossible to access this parameter directly. Instead, 
it follows from Equations \eqref{gcrf} and \eqref{gen_ll} that only the \emph{sum} $S_X^\ast - L_X^\ast$ can be inferred (similarly, only $S_{ZX}^\ast - L_{ZX}^\ast$ is accessible). The maximiser of the log-likelihood \eqref{gen_ll} is not unique and the problem is fundamentally misspecified.

We are therefore facing two related, but distinct, problems:
\begin{itemize}
\item \emph{identifiability}: under which conditions does the problem admit a \emph{unique} solution? 
Ideally, these conditions ought to be as broad as possible so that they will be met in realistic situations. Notice that unlike the breakdown caused by the high-dimensional regime, 
this kind of non-identifiability is more fundamental and remains no matter how large the number of samples.
\item \emph{consistency}: provided there exists a unique solution, can we derive
a consistent, tractable estimator which is capable of recovering $(S_X^\ast, L_{X}^\ast, S_{ZX}^\ast, L_{ZX}^\ast)$?
\end{itemize}

Here, we chose to focus on $S_X^\ast$ because it fits our application but there might be situations in which other parameters are of interest, \emph{e.g.} $S_{ZX}^\ast$ in \cite{Zhang2014}.

\subsection{Previous Work}

In practice, model selection in the context of GGMs is often performed
using $\ell_1$-regularised maximum likelihood estimators (MLEs) such as
the ones introduced in \cite{Banerjee:2008:MST:1390681.1390696, Yuan2007}, and the so-called \emph{graphical lasso} \citep{Friedman2008}. The $\ell_1$-norm is the convex envelope of
the $\ell_0$ unit ball and is therefore a natural convex relaxation to learn sparse
matrices. Building on the success of the graphical lasso, estimators
of the form ``log-likelihood'' + ``non-Euclidian convex penalty'' have received 
considerable interest \citep{Chandrasekaran2012a}. A relevant example is the use of the nuclear norm (\emph{i.e.} the sum of the singular values) as a convex relaxation for learning low-rank models \citep{Bach2008}. Beyond their attractive 
computational properties, the $\ell_1$ and nuclear norm regularised MLEs
enjoy strong theoretical guarantees \citep{Bach2008,Ravikumar2011}.

Using penalised MLEs, the questions raised above (Section \ref{goal}) have been solved in some special cases of model \eqref{gcrf}.


\subsubsection*{Sparse Gaussian Conditional Markov Random Field: $H = \emptyset$}

When $H$ is empty, \eqref{gcrf} reduces to
\begin{equation*}
    Y_{X} | Y_Z \sim \mathcal{N}\left\{
    -{S^\ast_X}^{-1}
    {S^\ast_{ZX}}^T Y_Z, 
    {S^\ast_X}^{-1} \right\}.
\end{equation*}
The log-likelihood associated with this model is convex and maximum-likelihood
estimates can be obtained in closed form. In order to increase the 
interpretability of the estimates and cope with high-dimensionality, \cite{Sohn2012, Wytock2013} suggested the following estimator of
$(S^\ast_{X}, S^\ast_{ZX})$:
\[
    (\hat{S}_X, \hat{S}_{ZX}) = \argmin_{S_X \in \mathbb{R}^{p \times p},
    S_{ZX} \in \mathbb{R}^{m \times p}, S_X \succ 0} -\ell(S_X, S_{ZX};\Sigma^n_Z, \Sigma^n_X)
    + \lambda_n (||S_X||_1 + ||S_{ZX}||_1),
\]
with $\lambda_n > 0$.  The entries of both $S_X$ and $S_{ZX}$ are being shrunk
in order to jointly learn a pair of sparse matrices describing the direct effects of $Z$ on $X$ and the graph over $X$. \cite{Wytock2013} studied the
theoretical properties of this estimator and derived a set of sufficient conditions for the correct recovery of $S^\ast_X$ and $S^\ast_{ZX}$. Among other results, they showed that this approach often outperforms the graphical lasso in terms of predictive power and model selection accuracy. Alternative parametrisations and approaches have been suggested in the multivariate linear regression literature. We refer the reader to \cite{Yin2011, Sohn2012} and references therein for more details on these estimators and their relative performances.


\subsubsection*{Low-Rank Plus Sparse Decomposition: $Z = \emptyset$}

The presence of latent variables ($H \neq \emptyset$) is a substantial
complication. As explained earlier, the marginal precision $S^\ast_X - L^\ast_X$ is then the sum of two matrices and the problem is fundamentally misspecified. However, following the seminal work of \cite{Candes2011} and \cite{Chandrasekaran2009}, \cite{Chandrasekaran2012} showed that it is 
sometimes possible to correctly decompose $S^\ast_X - L^\ast_X$ into its summands. Loosely speaking, this is the case if $S_X^\ast$ is sparse and there are few hidden variables with an effect spread over most of the observed variables. As a result, \cite{Chandrasekaran2012} introduced an estimator
which penalises the $\ell_1$-norm of $S_X$ and the nuclear norm of $L_X$
as follows:
\begin{equation}\label{Ch_obj_func}
    (\hat{S}_X, \hat{L}_X) = \argmin_{S_X, L_X \in \mathbb{R}^{p \times p}}
    - \ell(S_X,L_X;\Sigma_X) + \lambda_n (\gamma ||S_X||_1 + ||L_X||_\ast),
\end{equation}
subject to $S_X - L_X \succ 0, L_X \succeq 0$., with $\lambda_n, \gamma > 0$. Here $||L_X||_\ast$ denotes the nuclear norm of $L_X$ (\emph{i.e.} the sum of its singular values). Among other useful results, \cite{Chandrasekaran2012} showed that this estimator is, under suitable conditions, sparsistent and ``ranksistent'': the sign patterns of both the entries of $S$ and the spectrum of $L$ can be recovered exactly. 

\subsection{Suggested Estimator}

As hinted in the introduction, there are many cases where one might
want to both condition and allow for latent variables. In such cases, 
neither the sparse Gaussian conditional Markov random field nor the low-rank plus sparse approach would be optimal.
Building on these estimators, we propose decomposing the parameters of a Gaussian conditional Markov random field into the sum of a low-rank and a sparse matrix. To that end, we suggest optimising
the following regularised MLE
\begin{multline} \label{obj_func}
    (\hat{S}_X, \hat{L}_X, \hat{S}_{ZX}, \hat{L}_{ZX}) = \\\argmin_{S_X, L_X \in \mathbb{R}^{p \times p}; S_{ZX}, L_{ZX} \in \mathbb{R}^{m \times p}} -\ell(S_{X},L_{X},S_{ZX},L_{ZX}; \Sigma^n_Z,\Sigma^n_X,\Sigma^n_{ZX}) +
    \lambda_n (\gamma \norm{S}_1 + \norm{L}_\ast)\\
    \text{s.t.}~ S_X - L_X \succ 0, L_X \succeq 0
    ~\text{and}~ S = \begin{pmatrix}
     S_X  \\
     S_{ZX} 
\end{pmatrix},
    L = \begin{pmatrix}
     L_X  \\
     L_{ZX}
    \end{pmatrix}.
\end{multline}
Solving \eqref{obj_func} amounts to minimising a function which is \emph{jointly convex} in its parameters over
a convex constraint set (proofs are in the supplementary materials, along with other elementary properties of the likelihood).
As mentioned earlier, this likelihood is structured around two parameters, $S_{ZX}$ and $S_X$, accounting respectively for the direct (\emph{i.e.} conditional on all variables) effects of $Z$ on $X$
and the structure of the graph over $X$. However, because we penalise the rank of $L$, the effect of all latent variables is modelled jointly and a single set of latent factors is learned. No distinction is being made between the variables that ``mediate'' the
action of $Z$ and the ones that act as confounders on $X$. On the other hand, the parameters $S_X$ and $S_{ZX}$ retain their interpretability.
\section{Theoretical Analysis}
\label{theo_results}
According to our assumptions, we assume here that each sample is generated according to the model 
\begin{equation} \label{generative}
    Y_X | Y_Z \sim \mathcal{N}\left(-(S_X^\ast - L_{X}^\ast)^{-1} (S_{ZX}^\ast - L_{ZX}^\ast)^T Y_Z, \left(S_X^\ast - L_{X}^\ast \right)^{-1} \right),
\end{equation}
and ask under what circumstances Estimator \eqref{obj_func} correctly recovers the parameters $S^\ast, L^\ast$ (as built by stacking $S_{X}^\ast, S_{ZX}^\ast$ and  
$L_{X}^\ast, L_{ZX}^\ast$) with overwhelming probability.

We analyse this problem in the framework of \cite{Chandrasekaran2012} and therefore our proofs often mirror theirs. However, because of the form taken by the likelihood and because we do not limit
ourselves to square matrices, the analysis is significantly more involved.

As mentioned earlier, modelling latent variables by decomposing the parameters into a sum of two matrices raises \emph{identifiability} issues: given samples drawn from \eqref{generative}, when is it possible 
to exactly decompose the sum $S - L$ (where $S, L$ are defined as before) into its summands? This is a problem which
has been tackled in great generality in \cite{Chandrasekaran2012} and their results directly apply to the present situation: they are expressed in terms of the Fisher Information Matrix but do not explicitly involve the likelihood function. For that reason,
key definitions, as well as assumptions necessary for our result to hold, are deferred to the supplementary materials. Here we focus on the original contributions of this paper by
giving an intuition for these conditions before formally stating the \emph{consistency} of 
the estimator defined by \eqref{obj_func}.

\subsection{Identifiability}
\label{theo_results::identifiability}
Until now, it was repeatedly mentioned that a ``low-rank plus sparse decomposition''
is possible when $S$ is sparse and $L$ is low-rank. However, it is clear that imposing 
conditions on the sparsity of $S$ and the rank of $L$ is not sufficient. For example, consider 
a matrix with a single entry: it is at the same time sparse and low-rank and there is, therefore, no unique way of decomposing it into the sum of a low-rank and a sparse matrix. \cite{Chandrasekaran2009}
introduce the notion of \emph{rank-sparsity incoherence} and define quantities that make it
possible to express the conditions under which such a problem is well-posed, even for arbitrary matrices.
Two concepts are particularly important (precise mathematical statements and explanations
can be found in the supplementary materials):
\begin{itemize}
    \item $\xi(T(L^\ast))$: a small $\xi(T(L^\ast))$ guarantees that no single latent variable will have
a strong effect on only a small set of the observed variables. It is closely related to the concept of \emph{incoherence} introduced in \cite{Candes2011}. 
    \item $\mu(\Omega(S^\ast))$ quantifies the diffusivity of $S$'s spectrum. It can be shown that matrices with few non-zero entries per row/column (and thus sparse) have a small $\mu$.
\end{itemize}
A sufficient condition for identifiability can be expressed in terms of $\xi, \mu$ by requiring 
that their product be small enough ($\xi(T(L^\ast)) \mu(\Omega(S^\ast)) \leq \frac{1}{6}\mathcal{C}^2$) and that the tuning parameter $\gamma$ be chosen within a given range ($\gamma \in \left[\frac{3\xi(T(L^\ast))}{\mathcal{C}}, \frac{\mathcal{C}}{2\mu(\Omega(S^\ast))}\right]$),
for some constant $\mathcal{C}$ which depends on the Fisher Information Matrix (FIM). In other words, there must be a small number of latent variables acting on many observed ones and $S^\ast$ must not
have too many non-zero entries in any given row or column. This is a condition on the nominal parameters $S^\ast, L^\ast$ and it is related to the problem of decomposing the sum of two matrices. Moreover, it can be shown that
natural classes of matrices satisfy these assumptions. In particular, the degree of $S^\ast$ and number of latent variables $h$ are allowed to increase
as a function of the problem size $p,m$ \citep{Chandrasekaran2009}.
We call these restrictions on $\xi, \mu$ and $\gamma$ \textbf{Assumption 1}.

Another issue is that one does not directly observe $S^\ast - L^\ast$ but samples generated from \eqref{generative}. All lasso-type methods face this problem and conditions on the FIM are usually imposed (the so-called \emph{irrepresentability condition}) \citep{Ravikumar2011}. Similar assumptions about the FIM are made here and detailed in the supplementary materials. This is \textbf{Assumption 2}.

\subsection{Consistency}

We can now present our main result and state the consistency of Estimator \eqref{obj_func} (see supplementary materials for the proof). First, let us recall that for any matrix $P$, $||P||_2$ denotes its largest singular value and $||P||_\infty$ is its 
largest entry in magnitude. We can then define the following quantities: 
\begin{multline*}
\psi_Z = ||\Sigma_Z^n||_2, ~\psi^\ast_X = ||(S_X^\ast - L_X^\ast)^{-1}||_2, ~\phi^\ast_{ZX} = ||S_{ZX}^\ast - L_{ZX}^\ast||_2,\\ \psi = \frac{3}{2} \psi^\ast_X \sqrt{1 + 2 \frac{\psi_Z}{\psi^\ast_X} \left(1 + \frac{9}{4} \psi^\ast_X \phi^\ast_{ZX} \right)^2},
W = Q_1 \min\left(\frac{1}{6 \psi^\ast_X}, \frac{\phi_{ZX}^\ast}{4}, \frac{Q_2}{\psi^\ast_X \psi^2}\right).
\end{multline*}
Finally, for $M = \max \left(1, \frac{\psi_Z}{4\psi_X^\ast}(1 + \sqrt{\frac{m}{p}})^2\right)$, let $\lambda_n = \frac{Q_3}{\xi(T(L^\ast))} \sqrt{\frac{256 {\psi_X^\ast}^2 pM}{n}}$.
    
We prove the following theorem in the supplementary materials ($Q_1$ to $Q_6$ are constants
whose definitions are deferred for clarity):
\begin{theorem} \label{main_theorem} (Algebraic Consistency) \\
    Suppose that Assumptions 1 and 2 hold and that
    we are given $n$ samples drawn according to \eqref{generative}. Further assume that the following hold:
    \begin{enumerate}[(a)]
    \item $ n \geq \frac{pM}{\xi(T(L^\ast))^4} \max\left(2, \frac{256 {\psi_X^\ast}^2}{W^2} \right)$.
    \item ($\sigma_{\min}$ condition) Let the minimum non-zero singular value $\sigma$ of $L^\ast$ be such that
    \[
        \sigma \geq \frac{Q_4 \lambda_n}{\xi(T(L^\ast))^2}.
    \]
    \item ($\theta_{\min}$ condition) Let the minimum magnitude non-zero entry $\theta$ of $S^\ast$ be such that
    \[
        \theta \geq \frac{Q_5 \lambda_n}{\mu(\Omega(S^\ast))}.
    \]
    \end{enumerate}

    Then, with probability greater than $1 - \max\left(2 \exp(-pM), \exp(- 4\frac{\psi_X^\ast}{\psi_Z}pM) \right)$ we have
    \begin{enumerate}
    \item $\text{sign}(\hat{S}) = \text{sign}(S^\ast)$ and $\text{rank}(\hat{L}) = \text{rank}(L^\ast)$.
    \item 
    \[
        \max \left(\frac{1}{\gamma}\norm{\hat{S} - S^\ast}_\infty, \norm{\hat{L} - L^\ast}_2 \right)
         \leq \frac{Q_6 \psi_X^\ast}{\xi(T(L^\ast))} \sqrt{\frac{pM}{n}}.
    \]
    \end{enumerate}
\end{theorem}

Seen at a high-level, Theorem \ref{main_theorem} is analogous to the result obtained by \cite{Chandrasekaran2012} for the low-rank plus 
sparse (LR+S) estimator. In particular, this result holds when both
the dimension of the problem (parametrised by $m$ and $p$) and the number of samples ($n$) grow. Moreover, through their dependencies
on $\xi$ and $\mu$, both the degree of $S$ and the rank of $L$ are allowed to scale with $n$, which is essential to study connected graphs (see \citep{Chandrasekaran2009} for examples of scaling regimes). To compare these results further, a few points are worth considering.

First, we do not make any distributional assumptions about $Y_Z$ and there are therefore many scenarios in which only Theorem \ref{main_theorem} applies. 

For the sake of comparison, 
we can assume that $Y_Z$ follows a normal distribution so that the consistency theorem of LR+S is applicable.
Since LR+S does not model conditional distributions, $Z$ and $X$ are modelled jointly. The estimated matrices,
$(\hat{S}_{LR+S}, \hat{L}_{LR+S})$, are of size $(p+m) \times (p+m)$ and, to obtain $\hat{S}_X, \hat{S}_{ZX}, \ldots$, the relevant sub-matrices are extracted from the larger $(p+m) \times (p+m)$ estimates. 
Then, the algebraic consistency of LR+S holds with a probability $\pi_{LR+S}$ of at least $1 - 2\exp(-(p+m))$ provided the number of samples $n_{LR+S}$ satisfies $n_{LR+S} \gtrsim \frac{p + m}{\xi(T(L_{LR+S}^\ast))^4}$ \cite[Theorem 4.1]{Chandrasekaran2012}. Now, note that irrespective of the value of $M$, our convergence regimes are very similar since we 
require $n \gtrsim \frac{pM}{\xi(T(L^\ast))^4}$ for consistency to hold with a probability $\pi$ satisfying $\pi \geq 1 - c \exp(-pM)$, for some $c$. 
Since we have both $n_{LR+S} \neq n$ and $\pi_{LR+S} \neq \pi$, a direct comparison is not obvious. There are special cases in which it is easier. For example, when $M = 1$ one proves a result
similar to Corollary X (suppl. materials) showing that $n \gtrsim \frac{p +m}{\xi(T(L^\ast))^4}$ is required for $\pi$ to be at least 
$1 - 2 \exp(- (p+m))$, thus recovering a convergence rate similar to LR+S's.

Finally, $\mu$ and $\xi$ play an identical role in both Theorem \ref{main_theorem} and \cite[Theorem 4.1]{Chandrasekaran2012}, namely through \textbf{Assumption 1} and conditions a), b) and c). However, 
these quantities are usually different (\emph{i.e.} $\mu(\Omega(S^\ast)) \neq \mu(\Omega(S^\ast_{LR+S}))$,
$\xi(T(L^\ast)) \neq \xi(T(L^\ast_{LR+S}))$), which has interesting implications. An obvious consequence is
the one stated in the previous section: since $\mu, \xi$ define the acceptable range for $\gamma$, its span can vary widely across methods. More importantly, one shows that conditions a), b) and c) are driven by the lower-end of that range. Should it be assumed instead that $\gamma = \frac{\mathcal{C}}{2 \mu(\Omega(S^\ast))}$ (the upper-end), all three conditions would be relaxed \cite[Corollary 4.2]{Chandrasekaran2012}. Thus, 
the smaller the value of $\xi(T(L^\ast))$, the wider the acceptable range and the more likely Theorem \ref{main_theorem} is to hold.


\section{Optimisation}


Optimising \eqref{obj_func} in the high-dimensional setting is a challenging problem. For example, some of the constraints are hard to accommodate (\emph{e.g.} $S_X - L_X \succ 0, L_X \succeq 0$) and the penalty terms are non-smooth. Fortunately, \eqref{obj_func} has similarities with  
\eqref{Ch_obj_func} (the estimator of \cite{Chandrasekaran2012}) and we can rely
on algorithms that have proven effective on \eqref{Ch_obj_func}, namely
the Alternative Direction Method of Multipliers (ADMM) \citep{Boyd2010,Ma2013,Ye2011} and approaches relying on Semi-Definite Programming (SDP) \citep{Vandenberghe1996,Wang2010,Tutuncu2003}. The general theory behind both ADMM and SDP is applicable to the problem at hand but features that are specific to \eqref{obj_func} prevent a straightforward application of existing algorithms. SDP is an active field of research and recasting \eqref{obj_func} within that framework makes it easier for the reader to use existing software and even benefit from future advances in that field. On the other hand, our ADMM implementation is tailored to the problem at hand but converges to a reasonable accuracy quickly. This is why we discuss both strategies. Technical details and step-by-step derivations are given in the supplementary materials.


\subsection{The Alternating Direction Method of Multipliers}

The Alternating Direction Method of Multipliers (ADMM) is a first-order optimisation procedure which is well-suited to the minimisation of large-scale convex functions. 
It proceeds by decomposing the original problem into more amenable subproblems which are then solved iteratively \citep{Boyd2010}.
It is sometimes possible to obtain closed-form solutions to these subproblems but this is not required for ADMM to converge: even inexact iterative methods can be employed \citep{Eckstein1992, Goldstein2009}. 
Morover, only a few tens of iterations are necessary for ADMM to converge to an accuracy which is sufficient for most applications\footnote{However, converging to a very high accuracy can be slow in comparison to second-order methods.} \citep{Boyd2010}. ADMM (and related algorithms such as Bregman iterations and Douglas--Rachford splitting) has been celebrated as an efficient and robust general-purpose algorithm for $\ell_1$-regularised problems \citep{Goldstein2009}.

More recently, \citep{Ye2011, Ma2013} used ADMM to solve \eqref{Ch_obj_func} and showed that it can be optimised by iteratively solving four smaller subproblems \citep{Ye2011}. A similar decomposition is applicable to the problem at hand but, in the case of \eqref{obj_func}, one of the subproblems requires the computation of a so-called \emph{proximal operator} which does not admit a closed-form solution. Consequently, we derived an algorithm which iteratively converges to this proximal operator. In practice, we found that only a few iterations (typically less than 10) of this subprocedure are necessary to obtain a good approximation to the proximal operator. 


\subsection{Recasting the objective function as a Semi-Definite Program}

The solvers made available in the MATLAB\textsuperscript{\textregistered} packages SDPT3 and Logdet-PPA are capable of solving problems of the form \citep{Tutuncu2003,Wang2010}:
\begin{equation}
    \label{in:gen_prob}
    \argmin_{X_1, X_2, \ldots}~ Tr(X_1 C_1^T) + Tr(X_2 C_2^T) + \ldots + a_1 \log \det(X_1) 
\end{equation}
subject to a number of linear, quadratic and positive semi-definite constraints \footnote{This is only a subset
of the problems that can be tackled by such packages. See references for a formulation of this problem in its full generality.}. Our goal is then to recast \eqref{obj_func} as a problem of the same form as \eqref{in:gen_prob}.
We show in the supplementary materials that \eqref{obj_func} admits the following SDP reformulation:
\begin{multline}\label{in:sdp_prob}
    \argmin_{S_X, L_X, S_{ZX}, L_{ZX}, W, F, H_1, H_2} Tr(K\Sigma_O^n) - \log \det S_X + \lambda_n \left ( \gamma \mathbf{1}^T F \mathbf{1} + \frac{1}{2} \left(Tr(H_1) + Tr(H_2) \right) \right) \\
    \text{subject to}~ K\succeq 0,~ S_X \succ 0,~ L_X \succeq 0,~
    \left( \begin{array}{cc}
         H_1 & L\\
         L^T & H_2 
    \end{array} \right) \succeq 0,~
     -F_{ij} \leq S_{ij} \leq F_{ij},~\forall i,j;\\
     \text{where}~ K = \left( \begin{array}{cc}
         W & S_{ZX} - L_{ZX} \\
         S_{ZX}^T - L_{ZX}^T & S_X - L_X 
    \end{array} \right),~ S = \left( \begin{array}{c}
         S_X  \\
         S_{ZX} 
    \end{array} \right),~ L = \left( \begin{array}{c}
        L_{X}  \\
        L_{ZX} 
    \end{array} \right).
\end{multline}

\eqref{in:sdp_prob} can easily be implemented in \emph{e.g.} YALMIP and solved using LogdetPPA or SDPT3 \citep{Lofberg2004,Wang2010,Tutuncu2003}. We remark that the objective function is now smooth (as opposed to \eqref{obj_func}) but contains many more variables and constraints.
\section{Simulations}

We now study the properties of the proposed model on synthetic data and compare its performances
to the three other methods introduced earlier: the graphical lasso (GLASSO) \citep{Friedman2008}, the sparse conditional Gaussian graphical model (SCGGM) \citep{Sohn2012, Zhang2014, Wytock2013} and the low-rank plus sparse decomposition (LR+S) \citep{Chandrasekaran2012}. The suggested approach will henceforth be referred to as \emph{LSCGGM} (\emph{i.e.} Latent Sparse Conditional Gaussian Graphical Model). 

In Section \ref{theo_results},
it was established that assumptions about both the nominal parameters $(S^\ast, L^\ast)$ and the Fisher information matrix are necessary to guarantee the identifiability of the problem and, subsequently, the applicability of Theorem \ref{main_theorem}. In particular, we recalled the key role played by the maximum degree of $S^\ast$ and the incoherence of $L^\ast$. To better understand when these assumptions are expected to hold, we simulate data from a set of graphical models that span the range of possible latent structures and measure the ability of the different methods to recover the underlying graphs. 

\subsection{Graphical Structures and Methods}

The set of graphical structures we simulate from is constructed 
in such a way that only two integers, $d_Z$ and $d_H$, describe the relevant properties (rank, sparsity, incoherence, degree) of $S^\ast_{ZX} - L^\ast_{ZX}$ and $L_{X}^\ast$, respectively. Thus, $d_Z$ controls
the relationship between inputs $(Z)$ and outputs $(X)$ while $d_H$ encodes
the behaviour of $L_X^\ast$. The remaining parameter, $S_X^\ast$, remains unchanged throughout. We now briefly describe how the graphs are constructed but defer technical details to the supplementary materials ($\emph{e.g.}$ distribution of effect sizes). The code used to generate the data and fit our model is made available with this paper.

For all simulations, each observation is generated according to a model
of the form
\[
    \begin{pmatrix} Y_{X} \\
    Y_{H}
    \end{pmatrix} | Y_Z \sim \mathcal{N}\left\{-\begin{pmatrix} S^\ast_X& M^\ast_{XH}\\ {M^\ast_{XH}}^T & M^\ast_H \end{pmatrix}^{-1} 
    \begin{pmatrix} {M^{\ast}_{ZX}}^T\\ 0 \end{pmatrix} Y_Z, \begin{pmatrix} S^\ast_X& M^\ast_{XH}\\ {M^\ast_{XH}}^T & M^\ast_H \end{pmatrix}^{-1} \right\},
\]
with $Y_Z$ a random vector of size $p$ whose entries are drawn independently from a t-distribution with 4 degrees of freedom. $Y_X$ is also of size $p$. 
Here, $Y_X$ and $Y_H$ are drawn jointly from a conditional random Markov field but only $Y_X$ and $Y_Z$ are observed, which implies that $L_X^\ast = M^\ast_{XH} {M^\ast_H}^{-1} {M^\ast_{XH}}^{T}$. The matrices $S^\ast_X, L_X^\ast$ and ${M^{\ast}_{ZX}}$ are constructed as follows. 

The non-zero pattern of the $p \times p$ matrix $S_X^\ast$ is identical across all simulations and is similar to the one adopted by \cite{Wytock2013}: the graph over $X$ is a chain of $p$ variables in which one link out of five has been removed. The non-diagonal entries of $S^\ast_X$ are such that ${S^\ast_X}_{ij} \neq 0$, if and only if $i = j + 1$ and $i \not \equiv 0\ (\textrm{mod}\ 5)$.

As stated above, the rank/sparsity of $L_X^\ast$ is described by a single integer, $d_H$. 
Specifically, we assume that $p$ is an integer of the form $p = 2^k$ and pick $d_H \in \{0, 1, \ldots, k\}$. 
Then, for a fixed value of $d_H$, $M^\ast_H$ and $M^\ast_{ZX}$ are random matrices constructed so that: a) there are exactly $2^{d_H}$ confounders, \emph{i.e.} the rank of $L_X^\ast$ is $2^{d_H}$; b) each of the $2^{d_H}$ confounders impacts exactly $p / 2^{d_H}$ outputs; c) each output is connected to exactly one latent variable. Thus, when $d_H = k$, there is effectively no confounding since latent variables and outputs are in a one-to-one correspondence. When $d_H$ is much smaller than $k$, there are few confounders with an effect spread over many observed variables. When $d_H$ is set close to $k$, there are many hidden variables, each affecting a handful of outputs -- a gross violation of the identifiability assumptions.

Likewise, $d_Z$ accounts for the structure of $M_{ZX}^\ast$. Here again,
we assume $p = 2^k$ and pick $d_Z \in \{0, 1, \ldots, k\}$. Then, $M^\ast_{ZX}$ is designed to satisfy: a) $rk(M_{ZX}^\ast) = 2^{d_Z}$; b) each row/column of $M_{ZX}^\ast$ has exactly $p / 2^d_Z$ non-zero entries. 
The effect of $d_Z$ is easily interpreted. For example, $d_Z = k $ is an ideal situation where inputs and output are in a one-to-one correspondence. As $d_Z$ goes from $k$ to $0$, $M^\ast_{ZX}$ becomes denser and increasingly incoherent. When $d_Z$ is close to $k$, $M_{ZX}^\ast$ is estimated as a sparse matrix. When $d_Z$ is small, its decomposition is a single low-rank matrix.

Finally, since neither GLASSO nor LR+S model conditional distributions, we
use these estimators as described in Section \ref{theo_results}, \emph{i.e.} by first modelling $Z$ and $X$ jointly and then extracting submatrices of the estimates.

\subsection{Results}


In our simulations, we set $p = 32$, $n = 3000$ and let $(d_Z, d_H)$ take values in $\{2,3,4,5\}^2$. Each of these 16 designs is replicated 20 times, for a total of 320 distinct datasets. 

\begin{figure}
    \centering
    \includegraphics[width=0.85\textwidth]{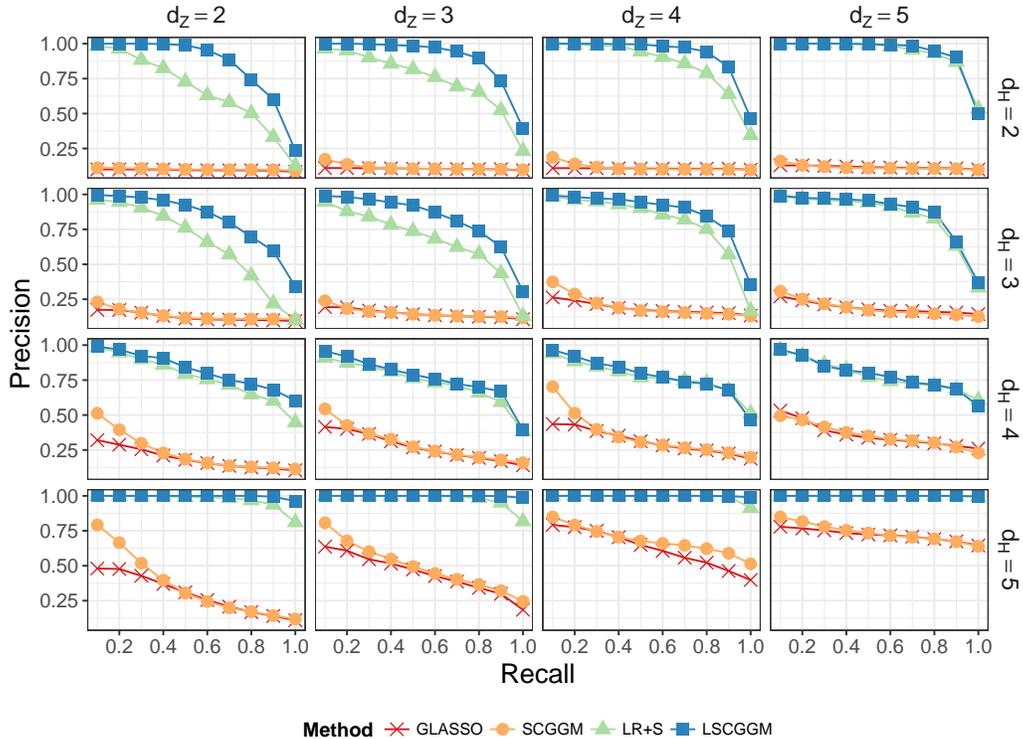}
    \caption{Comparison of the suggested estimator (LSCGGM) to other published methods.
    Along the $x$-axis (resp $y$-axis), $d_Z$ (resp $d_H$) varies from $2$ to $5$. More precisely,
    in the bottom row ($d_H=5$), there is no confounding at all. In the second row from the bottom ($d_H=4$), hidden variables
    act in a very sparse fashion. In the top row $(d_H=2)$, there are $4$ hidden variables and we are in the range of applicability of the low-rank plus sparse method. The second row ($d_H=3$) corresponds to an intermediate regime in which there are 8 latent variables. Settings: $p = 32$, $n=3000$. For each dataset, the value of the tuning parameter $\gamma$ was chosen so as to maximise the Area Under the Curve (AUC). Each of the 16 designs is repeated $20$ times. We report average precisions at fixed recalls of $\{0.1,0.2,\ldots,1\}$.}
    \label{sim1_SX_rocs}
\end{figure}

Here, we are interested in recovering the structure of $S^\ast_X$ and we use precision/recall curves as a metric, thus ignoring the rank of the latent component. LR+S and LSCGGM
both have two tuning parameters ($\lambda$ and $\gamma$). For each value of $\gamma$, one
obtains a distinct precision/recall curve by varying $\lambda$. For each of the 320 simulated datasets, we computed the paths corresponding to 15 distinct values of $\gamma$ and subsequently selected $\gamma$ so as to maximise the Area Under the Curve (AUC). Figure \ref{sim1_SX_rocs} shows the average precision/recall curves obtained by
applying this procedure.

First, we see that known methods behave as expected: GLASSO behaves best when there is no confounding and $Z$ acts in a sparse fashion ($d_H = d_Z = 5)$ ; SCGGM is more robust to changes in $d_Z$, but this is restricted to situations in which there is not confounding ($d_H =5$); LR+S performs best when $d_H = 5$ or when there is low-rank, diffuse confounding ($d_H = 2$). 
In a number of cases, the method proposed here is better than any of the alternative methods and, in the worst cases, it offers comparable performances. Specifically, it outperforms LR+S significantly when both inputs and hidden variables act on the outputs through a relatively low-rank mechanism ($d_Z = 2, 3; d_H = 2,3$). Two factors might explain this behaviour: a) the inputs are not normally distributed, which violates the assumptions of LR+S; b) the data is generated according to a conditional random Markov field, which is not assumed by LR+S, and may result in a violation of its identifiability assumptions.


$d_H = 4$ corresponds to the extreme situation in which each latent variable confounds exactly two random variables. None of the methods performs well but LR+S and LSCGGM behave better than GLASSO and SCGGM in scenarios where one would not expect any differences (\emph{e.g.} $d_Z = d_H = 5$). This is because LR+S and LSCGGM have \emph{two} tuning parameters, one of which ($\gamma$) is chosen with perfect knowledge:
it improves the AUC of these methods but causes $\hat{L}_X$ to be non-zero. 
Additional simulations made available in the supplementary materials show that when $\gamma$ is chosen with cross-validation, the selected value of $\gamma$ is indeed often too small.

Both LR+S and LSCGGM have two tuning parameters ($\lambda, \gamma$): 
$\lambda$ controls the overall shrinkage on the sparsity/rank of the estimates, $\gamma$ accounts for the trade-off between sparse and low-rank components. 
To better understand the role of $\gamma$, we look
at the precision/recall curves obtained for various values of this tuning parameter. As suggested in
\cite{Chandrasekaran2009}, the penalty term is reformulated as $\lambda (\gamma ||S||_1 + (1-\gamma)||L||_\ast)$ with $\gamma$ ranging from 0 to 1 instead of $(0, +\infty)$. 
By analogy to the AUC metric, we report the ``Volume Under the Surface'' (VUS) which accounts for the effect of both regularisation parameters.

\begin{figure}[b!]
    \centering
    \begin{center}
     \begin{subfigure}[t]{0.03\textwidth}
    \textbf{a)}
    \end{subfigure}
    \begin{subfigure}[t]{0.45\textwidth}
        \centering
        \adjustbox{valign=t}{\includegraphics[width=\textwidth]{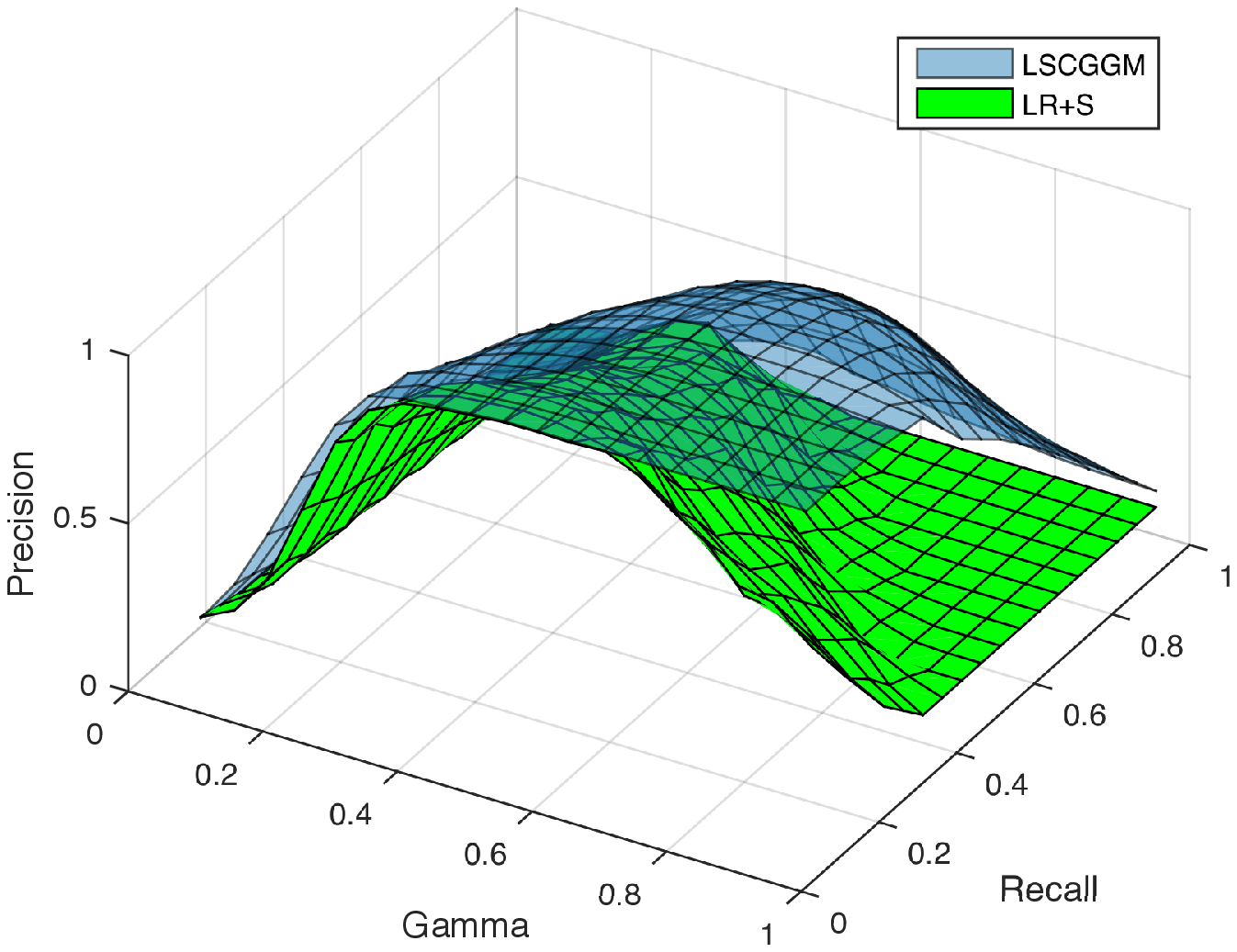}}
        \caption{}
    \end{subfigure}\hfill
     \begin{subfigure}[t]{0.03\textwidth}
    \textbf{b)}
    \end{subfigure}
    \begin{subfigure}[t]{0.45\textwidth}
        \centering
        \adjustbox{valign=t}{\includegraphics[width=\textwidth]{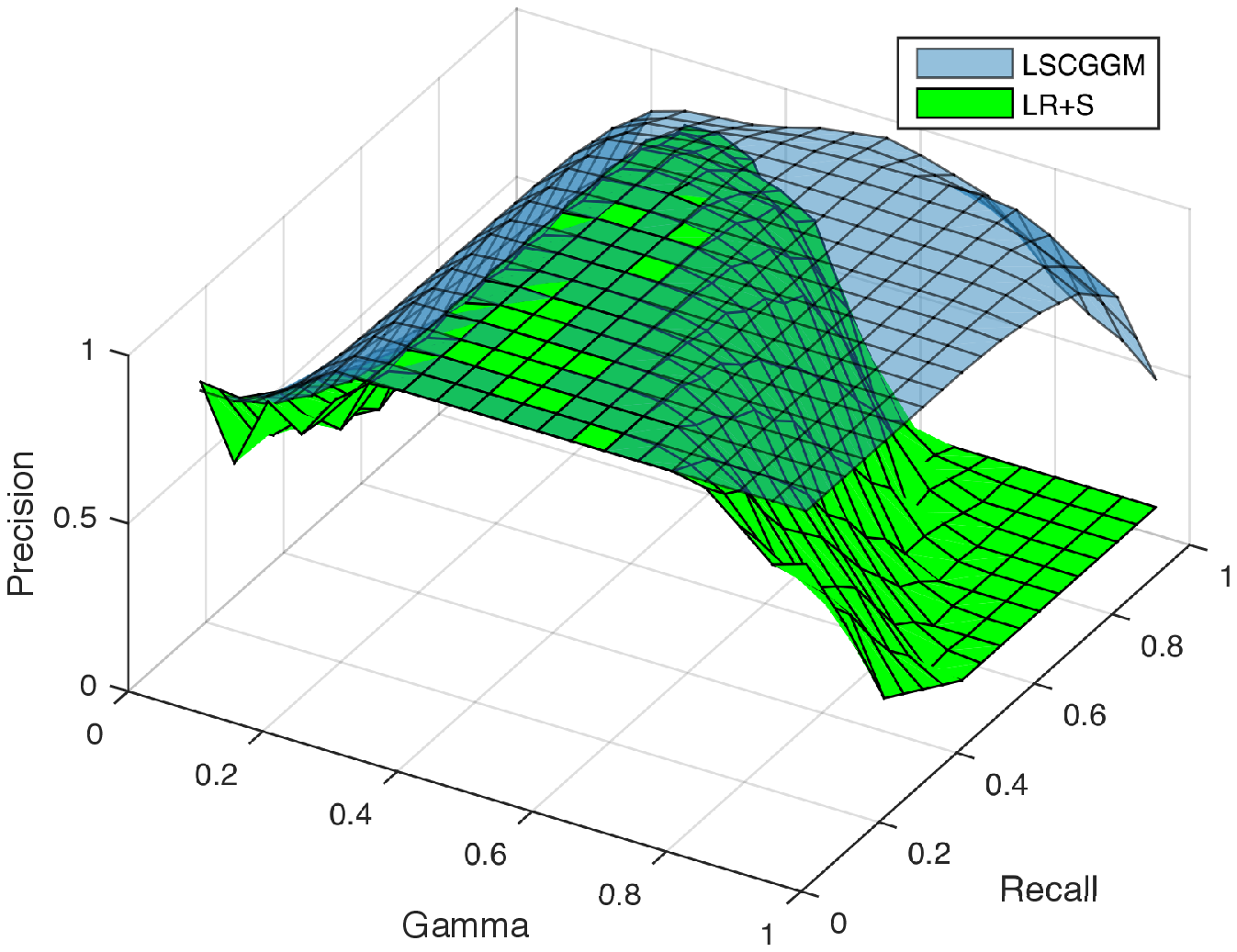}}
        \caption{}
    \end{subfigure} \\
     \begin{subfigure}[t]{0.03\textwidth}
    \textbf{c)}
    \end{subfigure}
    \begin{subfigure}[b]{0.74\textwidth}
        \adjustbox{valign=t}{\includegraphics[width=\textwidth]{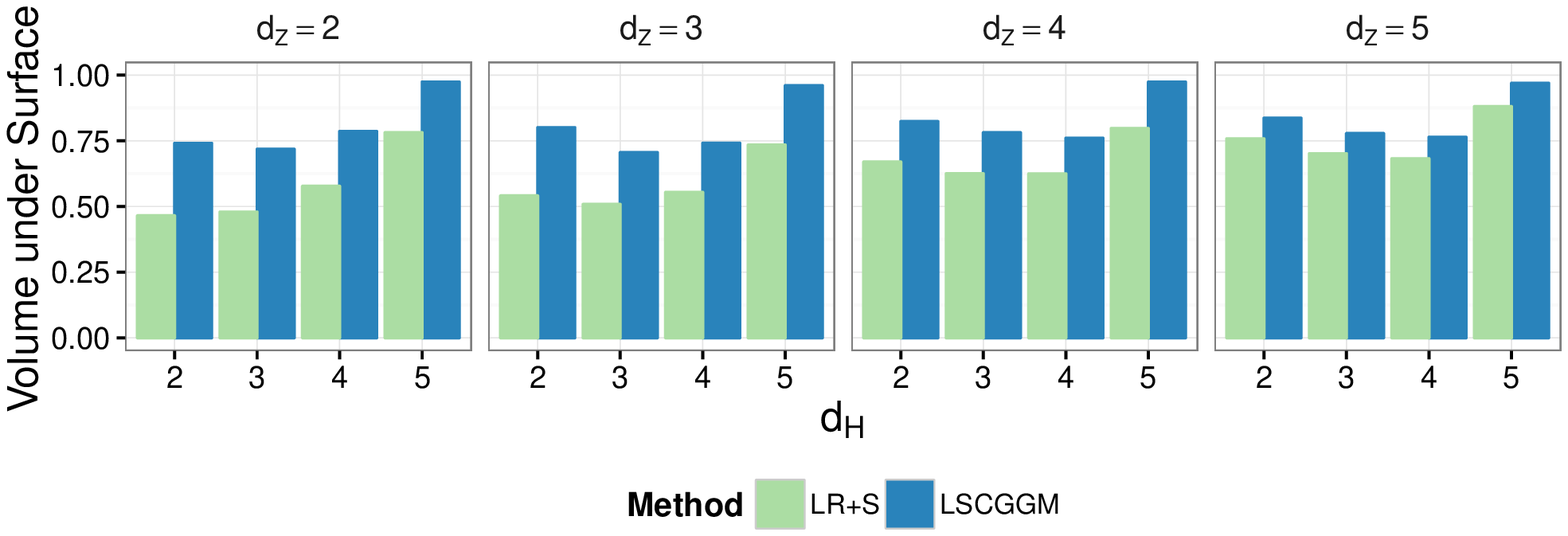}}
    \end{subfigure}
    \caption{Sensitivity to the tuning parameter $\gamma$. Here, an alternative
    parametrisation of the regularisation term is used: $\lambda_n (\gamma ||S||_1 + (1 - \gamma) ||L||_\ast)$, so that $\gamma \in (0,1)$ instead of $(0,+\infty)$.
	a) Precision/recall surface for $d_Z = d_H = 2$ (\emph{i.e.} each input acts on 8 random outputs and there are $4$ confounding variables).
	b) Precision/recall surface for $d_Z = 3$ and $d_H = 5$ (there are no confounders, each input acts on 4 random outputs).
	c) Volume under surface across all 16 simulation designs.
    }
    \label{sim1_VUS}
    \end{center}
\end{figure}

In Figure \ref{sim1_VUS}, the surfaces obtained for $(d_H = d_Z = 2)$ and $(d_H = 5, d_Z = 3)$ are plotted. 
They show that the suggested approach is less sensitive to $\gamma$ than LR+S, thus making it easier to pick a sensible value in real-world applications. Figure \ref{sim1_VUS} b) illustrates what happens when both methods offer comparable performances according to Figure \ref{sim1_SX_rocs} (which is obtained by choosing $\gamma$ \emph{perfectly}): compared to LSCGGM, there are actually very few values of $\gamma$ for which LR+S achieves its best AUC. Here, only two of the 16 possible surface plots are shown, but Figure \ref{sim1_VUS} c) indicates that LSCGGM is less sensitive to this tuning parameter across all simulation designs, as measured by the VUS. In particular, we have consistently observed that upper-end of the acceptable range for $\gamma$ is higher for LSCGGM than LR+S. The next simulations illustrate the implications of this property.


In these simulations, our main concern was to illustrate how methods differ in terms of identifiability and consistency. Setting $p$ and $m$ to a relatively small value (32) made it possible to capture most scenarios with only 16 graphical structures. 
In the supplementary materials, we simulate from larger graphs ($p = m = 2^7 = 128$, $n = 3000$) and obtain results that are similar to the ones showed here. 
We also report the estimation errors for the other parameters $L^\ast$, $L_{X}^\ast$ along with the precision/recall curves for $S_{ZX}^\ast$. Finally, we look at the effect of choosing $\gamma$ using cross-validation.
In the next section, we show how one can select $\lambda$ and $\gamma$ when some control over the number of falsely discovered edges is expected.

\section{Application: Using genetic information to detect relationships between human metabolites}
\label{Section:Application}

To illustrate the value of our new approach, we now apply it to a dataset combining human metabolite levels and genetic markers.
Here, metabolites play the role of the variables indexed by $X$ while genetic variants are the inputs, $Z$. 
For comparison purposes, we also report the results obtained with the Low-Rank plus Sparse method (LR+S)\footnote{The other two methods (graphical lasso and sparse conditional graphical model) 
arise as special cases by setting $\gamma$ close to $0$.}.

\subsection{The Avon Longitudinal Study of Parents and Children (ALSPAC)}

The Avon Longitudinal Study of Parents and Children (ALSPAC) is a cohort study of children born in the county of Avon during 1991 and 1992 \citep{Boyd2012,Fraser2012}. More details about this study and data preparation are available in the supplementary materials. Here, only key features of this dataset are reported\footnote{Please note that
the study website contains details of all the data that is available through a fully searchable data
dictionary (\url{http://www.bristol.ac.uk/alspac/researchers/access/}). Ethical approval for the study was obtained from the ALSPAC Ethics and Law Committee and
the Local Research Ethics Committees.}.

The data at our disposal contains genetic and phenotypic measurements on approximately 8,000 children and their mothers. We first performed our entire analysis on the children's cohort (called "Child cohort" throughout) and then independently applied the same procedure to the mothers' cohort (Mother cohort). We modelled the levels of 39 metabolites. Measurements
for all 39 variables were available without missing data for 5,242 children and 2,770 mothers. In each cohort, independent genetic variants were selected based on their predictive power with respect to any of the 39 traits under study: 133 and 44 variants were selected in the Child and Mother cohorts, respectively. Metabolite levels being continuous variables, they were quantile normalised and standardised. Genotypes, on the other hand, were encoded as ternary variables (0/1/2).

In summary, for the Child cohort (resp. Mother cohort) we have: $n = 5242,~ p = |X| = 39,~ m = |Z| = 133$ (resp. $n=2770,~p=39,~m=44$). 

\subsection{Methods}

Since both the suggested approach (LSCGGM) and the LR+S method have two tuning parameters ($\lambda, \gamma$), 
some procedure is required in order to set these parameters to appropriate values. As shown by both theoretical 
results and simulations, solutions are expected to be identical for a range of values of $\gamma$. Consequently,
we do not select a single value of $\gamma$ but consider instead $30$ values within the range $(0.02, 0.98)$\footnote{The penalty is parametrised as $\lambda (\gamma ||S||_1 + (1 - \gamma) ||L||_\ast)$, so that $\gamma \in (0,1)$.}. To each $\gamma$ corresponds a regularisation path: a graph along each path is
selected using ``pointwise'' complementary pairs stability selection \citep{Meinshausen2010,Shah2013}. Following the approach
used in \cite{Meinshausen2010}, the threshold on the inclusion probabilities is chosen by requiring 
that the expected number of falsely discovered edges be at most one: $E(V) \leq 1$ (using their notations).
Thus, for each method and each cohort we obtain a collection of thirty graphical structures.

In order to measure how similar two graphical structures are, we consider their edge sets. For any pair of undirected
graphs $\mathcal{G}_1 = (V_1, E_1),~ \mathcal{G}_2 = (V_2,E_2)$ we define their similarity by
their Jaccard Index
\[
    J(\mathcal{G}_1,\mathcal{G}_2) = \frac{|E_1 \cap E_2|}{|E_1 \cup E_2|}.
\]
This measure has two uses: 1) it makes it possible to select $\gamma$ by measuring how the estimates relate to each other
as $\gamma$ varies from $0$ to $1$; 2) it allows us to measure how well the findings are replicated across cohorts.

Another important step is assessing the biological relevance of the estimates using an external source of
information. We used ChEBI: an ontology of small chemical entities of biological interest \citep{Hastings2012}. 
We manually matched all 39 metabolites to their ChEBI IDs and annotated them using the ontology. Using such
annotations, one can compute an ``enrichment statistic'' reflecting whether a given graph contains edges between related metabolites
more often than would be expected in a random graph with a similar topology (such a graph has an expected statistic of $1$). We defer the definition of
this statistic to the supplementary materials but remark that this method is close to the ontology analyses frequently encountered in computational biology \citep{Wang2011}.
By randomly permuting the annotations, empirical p-values for this statistic can also be computed.

\subsection{Results}

\begin{figure}[b!]
    \centering
    \begin{subfigure}[t]{0.03\textwidth}
        \textbf{a)}
    \end{subfigure}
    \begin{subfigure}[t]{0.45\textwidth}
        \centering
        \adjustbox{valign=t}{\includegraphics[width=\textwidth]{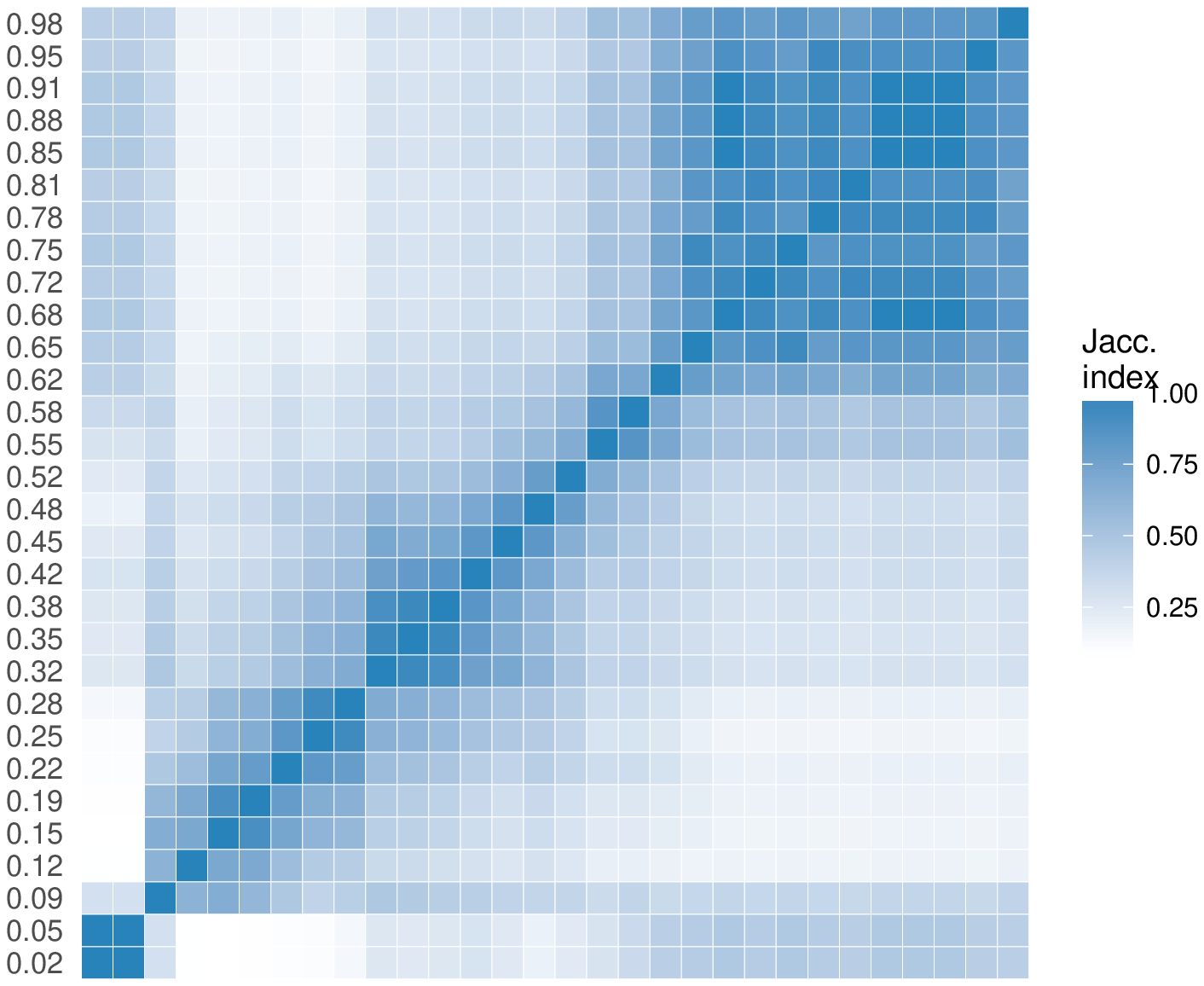}}
        \caption{}
        \label{gammaSens}
    \end{subfigure}\hfill
         \begin{subfigure}[t]{0.03\textwidth}
    \textbf{b)}
    \end{subfigure}
     \begin{subfigure}[t]{0.45\textwidth}
        \centering
        \adjustbox{valign=t}{\includegraphics[width=\textwidth]{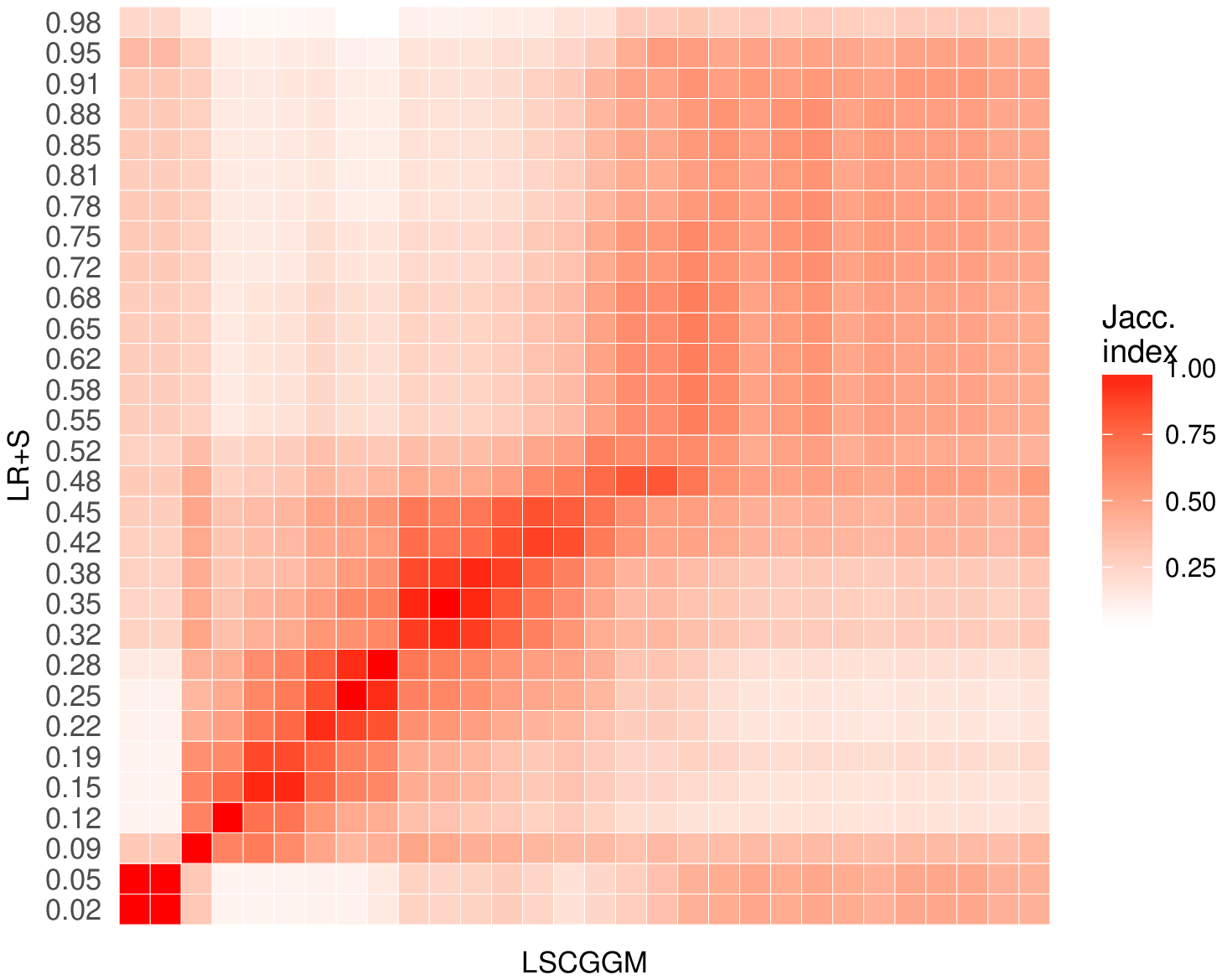}}
        \caption{}
        \label{compareGraphsMethods}
    \end{subfigure}
    \caption{Sensitivity of LSCGGM and LR+S to the tuning parameter $\gamma$. For any two graphs, their similarity is computed using the Jaccard Index of their edge sets.
	(a) Similarities between the edges sets of the graphs returned by LSCGGM in the Child cohort, as a function of $\gamma$ (for 30 values of $\gamma \in (0.02, 0.98)$).
	(b) Similarities between the graphs returned by LSCGGM and LR+S in the Child cohort.
	}
    \label{appResultsJacquard}
\end{figure}

First, we can ask how sensitive the estimates are to the tuning parameter $\gamma$. 
Indeed, as pointed out earlier, one would expect to see a ``stable region'': a range of values of $\gamma$ for which there is little variation. 
One would typically select a graph estimated with a $\gamma$ within this region. 
Let $\hat{\mathcal{G}}^{(\gamma)}_{LSCGGM,Ch}$ (resp. $\hat{\mathcal{G}}^{(\gamma)}_{LR+S,Ch}$) denote the graph returned by LSCGGM (resp. LR+S) for a given value of $\gamma$ in the Child cohort. 
For every pair $(\gamma_1, \gamma_2)$, Figure \ref{gammaSens} shows how similar the estimates are to each other (as computed by 
$J(\hat{\mathcal{G}}^{(\gamma_1)}_{LSCGGM,Ch},\hat{\mathcal{G}}^{(\gamma_2)}_{LSCGGM,Ch})$). In the range $0.6 \leq \gamma_1, \gamma_2 \leq 0.9$, they are very close
to each other. For small values of $\gamma$, the graphical structures returned by LSCGGM vary smoothly with $\gamma$. 
The regime $\gamma \leq 0.05$ corresponds to the case in which the rank of the latent component is $0$: LSCGGM behaves like a sparse conditional graphical model. Similar figures can be found for the LR+S method and the Mother cohort in the supplementary materials.

\begin{figure}[b!]
    \centering
        \centering
        \includegraphics[width=0.7\textwidth]{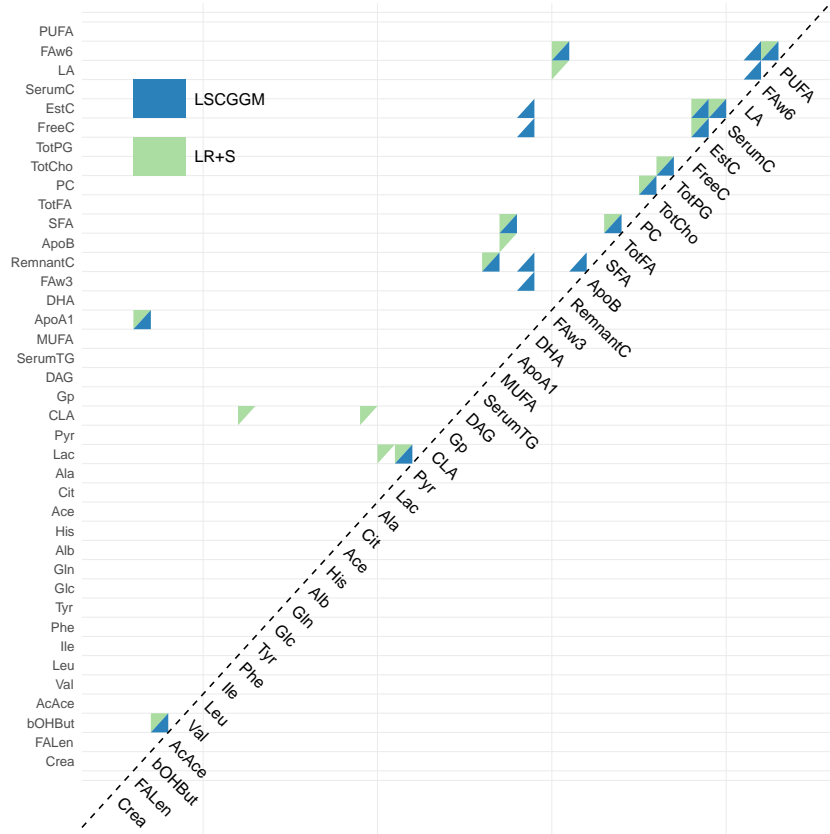}
    \caption{Adjacency matrices of the graphs returned by the LSCGGM and LR+S methods for $\gamma = 0.81$ and $\gamma = 0.68$ respectively.}
    \label{graphsComp}
\end{figure}

Having established that both methods exhibit a stable region, we look at how close
the estimates found in these regions are.
To that end, we plot $J(\hat{\mathcal{G}}^{(\gamma_1)}_{LSCGGM,Ch},\hat{\mathcal{G}}^{(\gamma_2)}_{LR+S,Ch})$
for all pairs $(\gamma_1, \gamma_2)$ (Figure \ref{compareGraphsMethods}). For small values of $\gamma$, LR+S and LSCGGM appear
indistinguishable.
However, for $\gamma_1, \gamma_2 > 0.5$ their Jaccard Index drops to reach values around
0.3 - 0.4. But the range $\gamma > 0.5$ covers precisely the stable regions of both LSCGGM and LR+S, thus indicating
that the methods' ``best guesses'' are different. 
Figure \ref{graphsComp} shows in what way the graphs found in those stable regions differ, with
LR+S inferring more connections between amino-acids. Here again, a similar result was obtained in the Mother cohort (see suppl. mat.). The supplementary materials also
contains the full name of the metabolites being modelled.

\begin{figure}[b!]
    \centering
     \begin{subfigure}[t]{0.03\textwidth}
        \textbf{a)}
    \end{subfigure}
    \begin{subfigure}[t]{0.70\textwidth}
        \centering
        \adjustbox{valign=t}{\includegraphics[width=\textwidth]{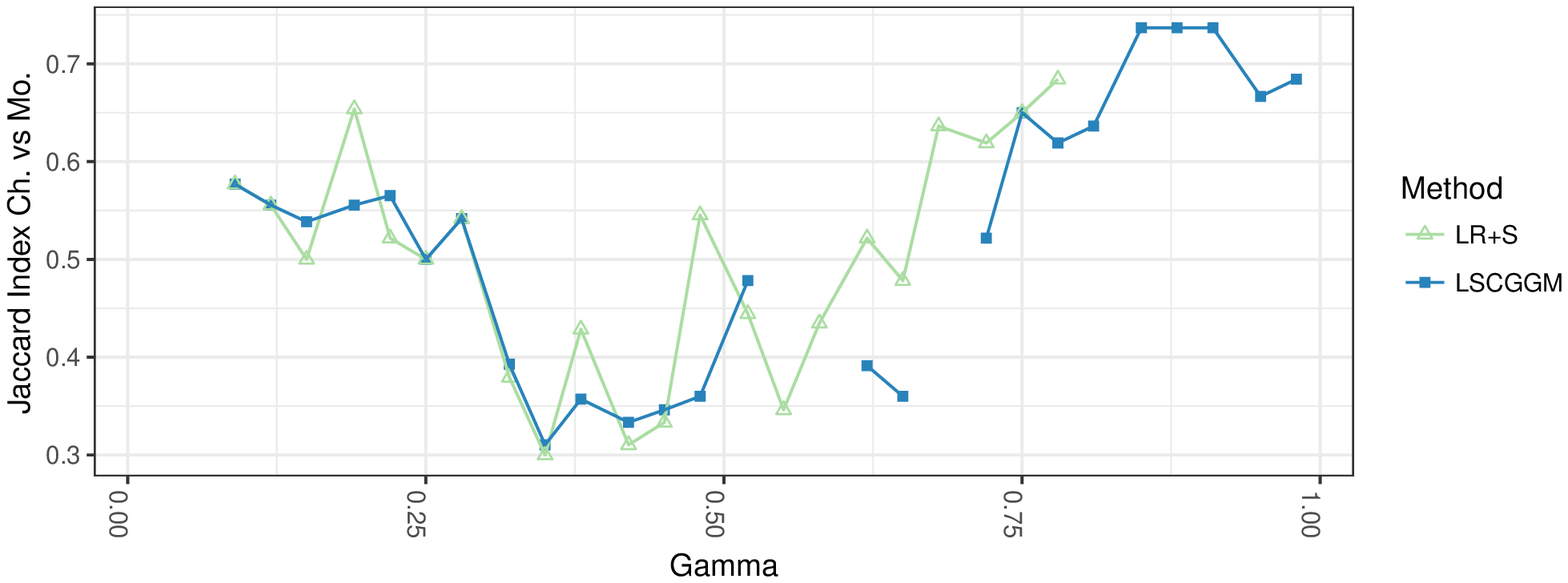}}
        \caption{}
	\label{repmetric}
    \end{subfigure}\\
     \begin{subfigure}[t]{0.03\textwidth}
        \textbf{b)}
    \end{subfigure}
    \begin{subfigure}[t]{0.70\textwidth}
        \centering
        \adjustbox{valign=t}{\includegraphics[width=\textwidth]{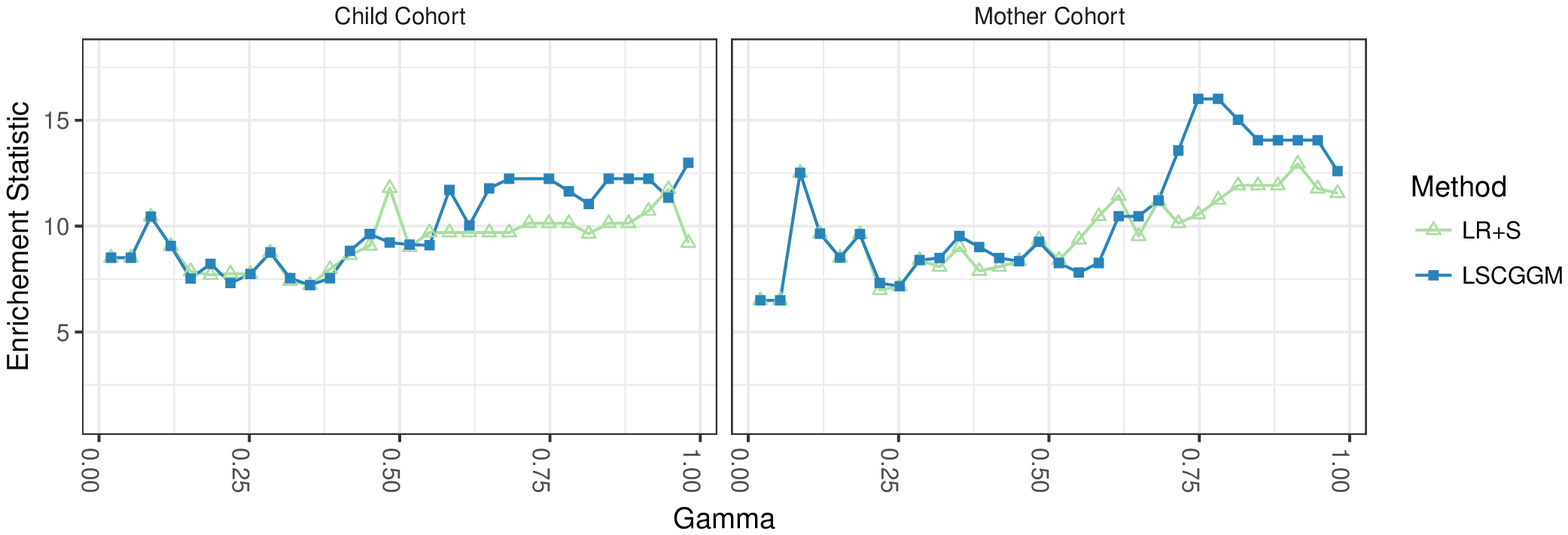}}
        \caption{}
	\label{enrichment}
    \end{subfigure}\\
    \caption{(a) Comparing estimates across cohorts. For each value of $\gamma$ and each method,
    we plot the similarity between the estimate obtained in one cohort against the one obtained
    in the other. We limit ourselves to values of $\gamma$ for which the estimates in both
    cohorts comport 15 edges or more. (b) Enrichment statistic, as a function of the tuning parameter $\gamma$.}
\end{figure}

Given that two cohorts are at our disposal, one way of assessing the 
quality of our results is to look at how well they replicate across datasets. In Figure \ref{repmetric} we plot
the similarity between graphs estimated at the same value of $\gamma$
 (see suppl. mat. for
 a plot of this similarity for all possible pairs $\gamma_1, \gamma_2$.). First, it can be seen
that higher replication values are achieved in the stable regions of their respective methods, with
Jaccard Indices at $0.6$ or above. We also see that LSCGGM's edge set replicates better
than LR+P's. Moreover, the suggested estimator retrieves more edges under the condition 
$E(V) \leq 1$ (see suppl. mat.).

Finally, we use the ``enrichment statistic'' defined earlier. In our attempt
to assess the quality of our estimates and their biological relevance, this
metric is useful as it makes it possible to score graphs using an external source of information.
Figure \ref{enrichment} shows the value taken by this statistic across cohorts and methods. Associated
p-values can be found in the see supplementary materials. Here again
it is clear that, irrespective of the dataset, higher values are achieved
within the stable regions of their respective methods. Just like in the case of the replication
measure, LSCGGM achieves the highest values. Given that the Child cohort
contains twice as many samples as the Mother cohort, it is surprising to observe
better performances in the Mother dataset. This might be due to the fact that this
cohort is more homogeneous: there are women only, measurements were taken the same number of months after pregnancy, etc...

\section{Discussion}

We discussed the problem of estimating a conditional Gaussian graphical model in the presence of latent
variables. Building on the framework introduced by the authors of \cite{Chandrasekaran2012}, we
suggested an estimator which decomposes the parameters of a sparse conditional Gaussian graphical 
model into the sum of a low-rank and a sparse matrix. Among other theoretical results, we established
that the proposed approach is well-behaved in the high-dimensional regime. Through simulations
and an application to a modern dataset comprising genetic and metabolic measurements, we compared
the performances of this approach to alternative methods.
In particular, we showed how such a conditional graphical model leads to better replication of the results across cohorts
and to estimates that are more biologically relevant.

The rise of high-throughput genetics, along with progress in data linkage, biobanking and functional genomics projects, has dramatically increased the number of datasets that include both genetic and multivariate phenotypic data. The data application we present in this paper, using genotype data to draw biological conclusions about the relationships between human traits, is thus becoming one of the most rapidly growing statistical challenges in human genetics. Conditional graphical models are particularly well-suited to such problems as they rely on an assumption we know to be true (namely, that genotype impacts phenotype and not vice-versa). Moreover, genetic
measurements are discrete in nature and it is therefore difficult to model them alongside continuous measurements. 
To the best of our knowledge, there are no approaches capable of learning a joint distribution over continuous and discrete
data in the presence of latent variables.

When it comes to lasso-type estimators, choosing an appropriate value of the tuning parameters can also be challenging. In simulations, our method seems to be less sensitive to the value of the tuning 
parameter $\gamma$, which makes it easier to set it to a suitable value in real life applications. Moreover, the
use of (complementary pairs) stability selection makes the estimates less sensitive to the value of $\lambda$
while providing some form of error control. 

Another limitation of such estimators comes from the fact that consistency/identifiability conditions are highly likely to be violated in real world applications. While this is true, a more realistic take is to regard our method 
as a means to generate ``causal'' hypotheses from a high-dimensional dataset. Paired with stability selection, such
an approach can realistically be used to generate a high-quality set of putative causal relationships that can then further
be investigated using hypothesis testing driven approaches (\emph{e.g.} instrumental variables). As shown in our application
this is an achievable goal.

Naturally, the method suggested here also suffers from a number of limitations and more work is required. For example, assuming that
the latent variables are normally distributed appears quite restrictive when compared to the flexibility offered by instrumental variable methods.
The question of learning discrete graphical models is also important but it is not yet clear how the present work
can be extended to such models.
\section*{Acknowledgements}

We are extremely grateful to all the families who took part in the ALSPAC study, the midwives for
their help in recruiting them, and the whole ALSPAC team, which includes interviewers,
computer and laboratory technicians, clerical workers, research scientists, volunteers,
managers, receptionists and nurses. The UK Medical Research Council and the Wellcome Trust
(Grant ref: 102215/2/13/2) and the University of Bristol provide core support for ALSPAC. This
publication is the work of the authors and Gil McVean will serve as guarantor for the
contents of this paper. GWAS data was generated by Sample Logistics and Genotyping Facilities at the Wellcome Trust Sanger Institute and LabCorp (Laboratory Corporation of America) using support from 23andMe.
This research was specifically funded by
the Wellcome Trust grant 100956/Z/13/Z (GM); the Wellcome Trust grant 098759/Z/12/Z (LJ)
and the EPSRC grant EP/F500394/1 and an Amazon Web Services research grant (BF).


\end{document}